\documentclass[a4paper]{article}
\usepackage{amsmath,graphicx}
\usepackage{multirow}
\usepackage{hyperref}
\newcommand{\masktoken}{{\fontfamily{qcr}\selectfont [MASK]}}

\usepackage{INTERSPEECH2021}

\title{PhonemeBERT: Joint Language Modelling of Phoneme Sequence and ASR Transcript}
\name{Mukuntha Narayanan Sundararaman$^{\dagger}$ \thanks{$^{\dagger}$ Work done during internship at Observe.AI} \qquad Ayush Kumar \qquad Jithendra Vepa}
\address{Observe.AI}
\email{mukunthas@observe.ai, ayush@observe.ai, jithendra@observe.ai}

\begin{document}

\maketitle
\begin{abstract}
Recent years have witnessed significant improvement in ASR systems to recognize spoken utterances. However, it is still a challenging task for noisy and out-of-domain data, where ASR errors are prevalent in the transcribed text. These errors significantly degrade the performance of downstream tasks such as intent and sentiment detection.
%In this work, we propose PhonemeBERT that learns a joint language model with phoneme sequence and ASR transcript to learn representations that are robust to ASR errors. 
In this work, we propose a BERT-style language model, referred to as \textbf{PhonemeBERT} that learns a joint language model with phoneme sequence and ASR transcript to learn phonetic-aware representations that are robust to ASR errors. 
% We also show that the pre-trained model can act as an encoder for noisy corpus requiring only ASR transcript as an input in low resource setups.
We show that PhonemeBERT leverages phoneme sequences as additional features that outperform word-only models on downstream tasks. We evaluate our approach extensively by generating noisy data for three benchmark datasets - Stanford Sentiment Treebank, TREC and ATIS for sentiment, question and intent classification tasks respectively in addition to a real-life sentiment dataset. The results of the proposed approach beats the state-of-the-art baselines comprehensively on each dataset. Additionally, we show that PhonemeBERT can also be utilized as a pre-trained encoder in a low-resource setup where we only have ASR-transcripts for the downstream tasks.
\end{abstract}
\section{Introduction}
\label{sec:intro}

%% CAN BE SHORTEN, CONTACT CENTER CONTEXT IS NOT REQUIRED, CAN WE REDUCE THIS ADD COUPLE OF LINES ON BERT
% With the proliferation of voice-enabled technologies, the globe has seen an upsurge in virtual assistants like Google Home, Alexa and Siri in daily usage. In addition to these devices, the legacy businesses revolving around voice calls in the call centers and BPO are at the cusp of disruption owing to the advancement in the speech and NLP technologies. 
With the proliferation of voice-enabled technologies
% in industries and devices, 
spoken language understanding (SLU) has become significantly ubiquitous.
%to understand the speaker utterances. 
%Such systems require spoken language understanding (SLU) to understand the speaker utterances to carry out the required tasks.
The general modus-operandi of SLU systems is to convert voice into text using an ASR engine and use Natural Language Understanding (NLU) on the transcribed text to comprehend the speaker’s intents and requests. Despite advancements in ASR systems, domain adaptation and word recognition in noisy setups remain a big challenge. 
% Proper nouns, out of vocab words and phonetically confusing words are often mistranscribed to common words. 
In a typical SLU system that operates on the ASR outputs, these errors degrade the performance of the system on the downstream tasks \cite{DBLP:conf/asru/DesotPV19}.

The approaches tried by scientific community to address the errors in the ASR system can be broadly categorized into four groups: a) \textit{Modelling word confidence}: Liu et al. \cite{liu2020jointlyArxiv} proposes a BERT model that jointly encodes the word confidence network and the dialog context. Ladhak et al. \cite{DBLP:conf/interspeech/LadhakGDMRH16} proposes LatticeRNN to encode the ambiguities of the ASR recognition for the intent classification task; b) \textit{ASR correction}: Weng et al. \cite{DBLP:conf/icassp/WengMKWZMNPWBT20} presents a contextual language correction on ASR outputs jointly with modelling LU task that learns from ASR n-best transcriptions. Mani et al. \cite{mani2020towards} use machine translation technique for domain adaptation to correct ASR mistakes in medical conversations; c) \textit{End-to-End SLU}: Serdyuk et al. \cite{DBLP:conf/icassp/SerdyukWFKLB18} explore the possibility to extend the end-to-end ASR learning to include NLU component and optimize the whole system for SLU task while Ghannay et al. \cite{DBLP:conf/slt/GhannayCECSLM18} study end-to-end named entity and semantic concept extraction from speech to circumvent the errors arising from the ASR pipeline; d) \textit{Phoneme enhanced representations}:  Yenigalla et al. \cite{DBLP:conf/interspeech/YenigallaKTSKV18} use word2vec generated phoneme embedding for emotion recognition task. Fang et al. \cite{DBLP:conf/sigir/FangFLR20} propose word2vec based approaches to learn phoneme embeddings capturing pronunciation similarities of phonemes to make classification robust to ASR errors.

\begin{figure*}[ht]
    \centering
    % \includesvg[width=1.0\textwidth]{emnlp_1.svg}
    % \includegraphics[width=0.95\textwidth,, height=6cm]{images/PhonemeBERT Architecture Diagram.pdf}
    % \includegraphics[width=0.95\textwidth]{images/PhonemeBERT Architecture Diagram.pdf}
    \includegraphics[width=0.95\textwidth, height=6cm]{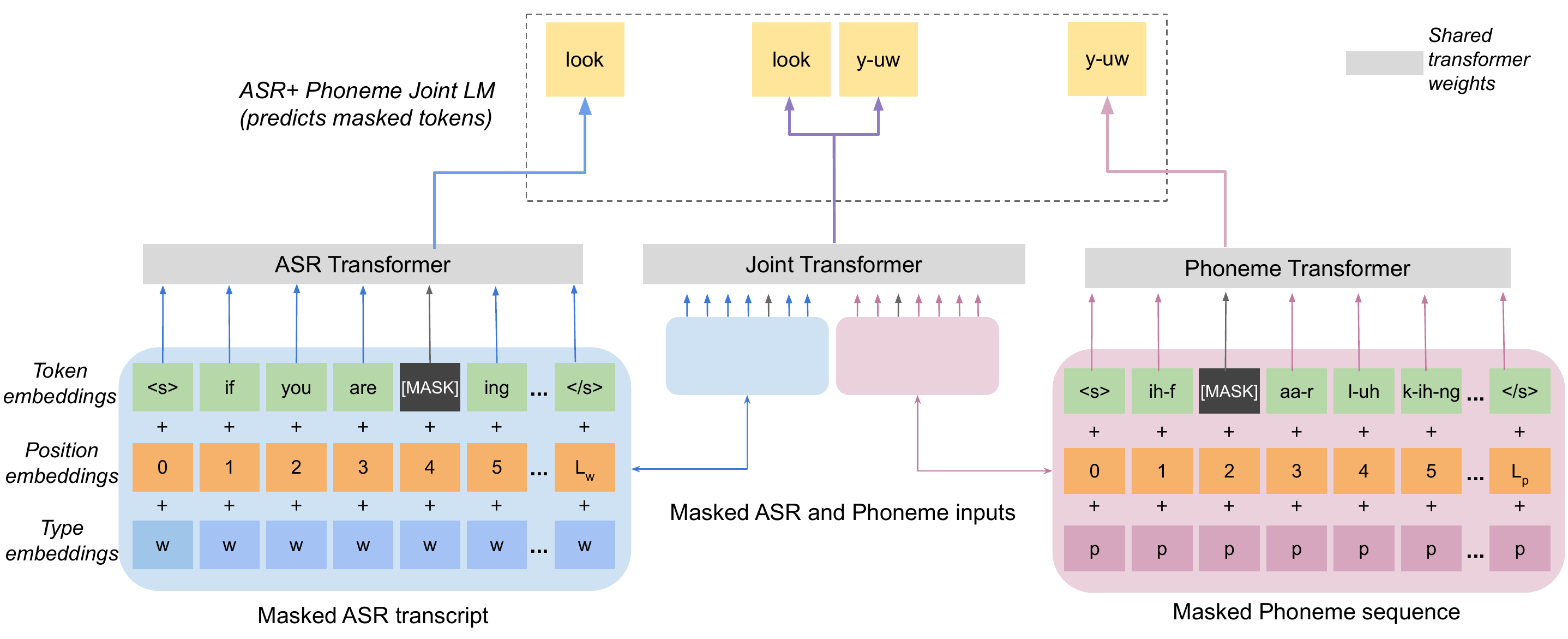}
    \caption{Proposed architecture of PhonemeBERT that jointly models phoneme sequence and ASR transcript.}
    \label{fig:architecture}
\end{figure*}

%\textit{[JITHENDRA'S INPUT] - Importance/utilization of phoneme} Our work falls in the last category of utilizing phonemes to learn representations that are more robust to ASR errors. However previous works disregard ASR transcript while training phoneme representations [PHRASE IT BETTER]. The authors use a word2vec model to train phoneme embeddings which does not take the information from the ASR sequence into the account. Furthermore, Fang et al.
Our work falls in the category of \textit{Phoneme enhanced representations} to learn representations that are robust to ASR transcription errors. %The phoneme information can be complementary to ASR transcript especially when the speech is more noisy.  To learn representations, 
In one of the earlier works, Yenigalla et al. \cite{DBLP:conf/interspeech/YenigallaKTSKV18} trained word2vec based phoneme and ASR embeddings independent of each other for SLU tasks.  
%either of these two inputs, phoneme sequences or ASR transcripts, 
% These methods do explore to model the complementary information present in them.
Fang et al. \cite{DBLP:conf/sigir/FangFLR20} propose multiple variants of phoneme embedding which either require alignments between the reference transcript and the phoneme sequence (\textit{p2va}) or learns phoneme embedding in isolation to the ASR transcript (\textit{p2vc, s2s}). Additionally, the authors derive phoneme sequence obtained on top of the ASR transcripts. We note two main drawbacks in these methods: \textit{a)} Learning embeddings from isolated sequences prohibits leveraging the complementary information present in the ASR transcript and phoneme sequence; \textit{b)} Due to direct conversion of words in ASR transcript to phoneme sequences, the errors from ASR transcript are propagated to the phoneme sequence which induces redundant errors that can lead to sub-optimal results.
% Instead, we
% We, instead, propose PhonemeBERT using phoneme sequences extracted directly from audio and learn representations in a pre-training methodology that jointly models the phoneme and ASR sequence.

% \begin{table}[]
% \small
% \centering
% \renewcommand{\arraystretch}{1.3}
% \caption{Example of a phoneme sequence extracted directly from audio (Ph-Speech) and on top of ASR output (Ph-ASR).}
% \begin{tabular}{l}
% % \hline
% % \multicolumn{1}{c}{\textbf{Examples}}                                                          \\ \hline
% \hline
% \textbf{Ref:} Why in tennis are zero points called love ?   \\
% \textbf{Hyp:} ten are zero points car love \textit{(WER: )}  \\
% \begin{tabular}[c]{@{}l@{}}\textbf{Ph-Speech:} W.AY.IH.N. .T.EH.N.AH.S. .AA.R. .Z.IY.R.OW. \\  .P.OY.N.T.S. .K.AO.L.D. .L.OW.V. \textit{(PER: )}\end{tabular} \\
% \begin{tabular}[c]{@{}l@{}}\textbf{Ph-ASR:} .T.EH.N. .AA.R. .Z.IH.R.OW. .P.OY.N.T.S. \\  .K.AA.R. .L.AH.V.  \textit{(PER: )}\end{tabular}
% \\\hline
% % \textit{Gold Truth}: DESC, \textit{PhonemeBERT}: DESC, \textit{B2}: ENTY                                           
% % Ref: A deliciously mordant , bitter black comedy .                       \\
% % Hyp: a published morgan than bitter black comedy                            \\
% % \begin{tabular}[c]{@{}l@{}}Ph: AY. .G.IH.L.IH.SH.AH.S.L.IY. .M.AO.R. .DH.AE.N.  \\    .S.IH.T.ER. .B.L.AE.K. .K.AA.M.AH.D.IY.\end{tabular} \\
% % Actual: Positive, PhonemeBERT: Positive, B3: Negative     
% \end{tabular}
% \label{qualitative}
% \end{table}

To address these drawbacks, we propose PhonemeBERT that jointly models the phoneme and ASR sequence with an added benefit of producing better results in a low-resource setup.
% The key contributions of this work are: 1) Joint modelling of phoneme sequence and ASR transcripts using a BERT based pre-training setup,
% %2) The dataset used in this work is released for future research work,
%     2) Extensive experiments are carried out on benchmark datasets to demonstrate the effectiveness of the approach, and, 
% %3) Proposed approach does not require alignment between ASR and phoneme sequences.
% %%CAN WE ALSO HIGHLIGHT TAKING ADVANTAGE OF PRE-TRAINED MODEL IF ASR TRANSCRIPT IS ONLY AVAILABLE
The main contributions of this work are: 
\begin{enumerate}
    \item PhonemeBERT: A method to jointly model ASR transcripts and phoneme sequences using a BERT-based pre-training setup is proposed.
    \item Extensive experiments are carried out on benchmark  and real life dataset to show the method's effectiveness. Results show that joint language model in PhonemeBERT can leverage phoneme sequences as complementary features, making it robust to ASR errors.
    \item Pre-trained PhonemeBERT can be effectively used as word-only encoder in a low-resource downstream setup where phoneme sequences are not available, still producing better results than word-only language model.
    \item We also release our generated datasets used in the work for research usages: \url{https://github.com/Observeai-Research/Phoneme-BERT} 
    % \item The proposed method does not rely on alignment between ASR transcript and phoneme sequence.
    
\end{enumerate}

% 1) ; 2) ;
%3) Experiments are carried out to show that PhonemeBERT also learns phonetic-aware representations that perform well in the absence of phoneme information during downstream task fine-tuning, where only noisy ASR transcripts are available.
% 3) .

\section{Proposed Methodology}
\label{method}

In the proposed work, we aim to learn representations that are robust to ASR errors. To accomplish this objective, we utilize the phoneme sequence in addition to the words (ASR transcript) to train a joint language model (LM), named \textit{PhonemeBERT} (Figure \ref{fig:architecture}). The vanilla masked language modelling (MLM) setup is trained by predicting the masked tokens using the contextual information from the surrounding tokens. We extend this setup to a joint MLM training that predicts masked tokens in both ASR transcript and the phoneme sequence leveraging the coupled information from the two sequences. In a joint MLM setup, we use a joint transformer to feed both ASR transcript and phoneme sequence to predict the masked tokens. To predict a token masked in the word sequence, the model can either attend to surrounding word tokens or to the phoneme sequence, encouraging the model to align the word and phoneme representations, making the word representations more phonetic-aware. The joint modelling of word and phoneme helps the model to leverage the phoneme context if the word context is insufficient to infer the masked token in the word sequence and vice-versa. In addition to joint loss, we also use shared transformers for ASR, phoneme and combined sequence represented by ASR transformer, phoneme transformer and joint transformer respectively in Figure \ref{fig:architecture}. Shared transformers ensures that shared representations are learnt that facilitate predicting masked token with joint as well as individual contexts while not adding complexity to the overall architecture.
% This step ensures that phonemes are leveraged while predicting masked tokens in the word sequence.
The proposed joint setup is optimized with a training objective comprising three loss functions: ASR MLM loss, phoneme MLM loss and joint MLM loss as defined in Eqn \ref{mlm}:.
% L_MLM, L_joint
% alpha, beta, gamma = 1 in our current paper
% $\alpha, \beta, \gamma$ represents the weights given to each MLM task to train PhonemeBERT.
\begin{equation}
\begin{split}
    \mathcal{L} = \sum_{i=1}^{|A_{mask}|}\mathcal{L}_{mlm}(a_i|\hat{A}) + \sum_{j=1}^{|P_{mask}|}\mathcal{L}_{mlm}(p_j|\hat{P}) \\ + \sum_{k=1}^{|A_{mask} + P_{mask}|}\mathcal{L}_{joint}(t_k|\hat{A},\hat{P})
\end{split}
    \label{mlm}
\end{equation}
where, $\mathcal{L}(t_i|\hat{T})$ represents the MLM loss for predicting token $i$ in using information from masked sequence $\hat{T}$ $\in (A, P)$ representing masked ASR transcript and phoneme sequence respectively. The loss is accumulated over all masked tokens ($A_{mask}, P_{mask}$) across both ASR transcript and phoneme sequence.
% \begin{equation}
%     L = L_1(a_i|A) + L_2(p_j|P) + L_3(a_i|A,P) + L_4(p_j|A,P)
%     \label{mlm}
% \end{equation}

% Instead of using individual MLM tasks for word and phoneme sequences, we propose a joint modelling by training the model on a parallel corpus of words and corresponding phoneme sequences. 

% QUESTION -- What is PhonemeBERT initialized with for phonemeBPE?

To train PhonemeBERT, the model is initialized with weights from RoBERTa \cite{DBLP:journals/corr/abs-1907-11692}. RoBERTa is a robustly optimized pre-trained LM using masked language modeling (MLM) as the pre-training objective.  % that also operates on phoneme byte-pair encodings (BPE). 
The training sequence in the proposed model contains the ASR transcript concatenated with the phoneme sequence (Figure \ref{fig:architecture}). 
We utilize Byte Pair Encoding (BPE) that represents the vocabulary of words and phonemes. BPE has been shown to be an effective method to handle large vocabularies with sub-word units. It has also been shown to work well with sub-words based on phonemes \cite{zeyer2020investigations}. We use a byte-level BPE vocabulary like in RoBERTa \cite{DBLP:journals/corr/abs-1907-11692}, which  allows using spaces (pauses in the phoneme sequence) as part of the BPE tokens. To represent phoneme sequences, we train our model with a vocabulary consisting of 600 phoneme sub-word units constructed over the phoneme sequences from the pre-training corpus.
% , 
% We randomly initialize the phoneme-BPE token embeddings.
Finally, we represent each BPE-token using three embeddings: \textit{token, position} and \textit{type embeddings}. Token embeddings for word sequence are initialized from RoBERTa base model, while phoneme-BPE token embeddings are randomly initialization and trained. Position embedding is initialized from RoBERTa base model. The position embeddings for the phoneme sequence start at 0 to enable soft yet unsupervised alignment between word and phoneme tokens. Following the training regime of RoBERTa, we sample randomly 15\% of the BPE tokens from the word and phoneme sequences and replace them by \masktoken. Of the selected tokens, 80\% are replaced with \masktoken, 10\% are left unchanged, and 10\% are replaced by a randomly selected vocabulary token of the same type (word or phoneme). In addition to token and position embedding, we randomly initialize a \textit{type embedding} denoting whether the token is coming from a `word' or a `phoneme' sequence.

\begin{table}[]
\footnotesize
\centering
\renewcommand{\arraystretch}{1.2}
\caption{Statistics of pre-training dataset}
\begin{tabular}{lcll}
Dataset & \#Sentences & \begin{tabular}[c]{@{}l@{}}Speech \\ Hours\end{tabular} & \begin{tabular}[c]{@{}l@{}}Vocab \\ Size\end{tabular} \\ \hline
LibriSpeech   & 35411 & 360\textit{h}      & 37468                                                                                                     \\
AmazonReviews & 75000  & 684\textit{h}      & 99583                                                                                                    \\
SQuAD v1.1        & 50000    & 278\textit{h}    & 134287            
\end{tabular}
\label{dataset_pt}
\end{table}

\subsection{Pre-training Dataset}
With a primary objective to learn generic representations for the noisy data, we choose three datasets from different domains to accomplish the pre-training step (Table \ref{dataset_pt}). We use, LibriSpeech corpus \cite{DBLP:conf/icassp/PanayotovCPK15}, Amazon reviews \cite{DBLP:conf/www/HeM16} and Squad v1.1 \cite{DBLP:conf/emnlp/RajpurkarZLL16} for pre-training step. Specifically, we use LibriSpeech train-clean-360 corpus, while we randomly sample the Amazon reviews and SQuAD dataset to collect specified number of sentences across the multiple topics and paragraphs. 
Once we collect the raw text corpus, we follow a similar strategy as described in Fang et al. \cite{DBLP:conf/sigir/FangFLR20} to create noisy speech corpus. We use Amazon Polly 
\footnote{\url{https://aws.amazon.com/polly}}
% \footnote{ \url{https://https://pair-code.github.io/facets/}}
to convert the raw text to speech. Next, we apply Speech Synthesis Markup Language (SSML) tags to the audio to change the prosody of the produced speech. We also add ambient noise 
\footnote{\url{www.pacdv.com/sounds/ambience_sounds.html}} 
to the speech to make the data align with real-life data. For a specific speech file, we apply one SSML tag and one ambient noise. Once the noisy speech data is generated, we use Amazon Transcribe to transcribe the audio. 
Since our hypothesis is to create a language model that is robust to ASR errors, we generate data spanning across different levels of word-error-rate (WER). 
%Based on our empirical analysis, we note that any transcription with lower WER is very close to the ground truth and with very high noise, the transcription becomes incomprehensible. 
% In this work, we discard all transcriptions that has word-error-rate (WER) lower than 5\%. Furthermore, with very high noise, the audio itself becomes incomprehensible and so is the generated ASR transcript. Based on empirical analysis,
% Hence to achieve a balance between the two extremes, 
We choose to reject any transcription that has WER less than 5\% or more than 40\%. We additionally use more than one noise level for 25\% of the pre-training corpus to capture characteristics of word and phoneme sequences at different noise levels. The mean WER for the generated corpus is 30.79\%.

% explicitly induce word and phoneme confusions/correlations (or correlation or something else?) in the dataset \textit{[CHECK WITH JITHENDRA]}. The mean WER for the generated corpus is 31.2\%.

% // we can capture characteristics of word and phoneme seq. at different noise levels.

%with the lowest WER for an utterance at 16\% and 

%\textit{[JITHENDRA'S INPUT]} In addition to transcribing the audio/speech, we generate phoneme sequence from a separate model trained using acoustic modelling. The primary objective to use an independent acoustically trained phoneme-generator is make it non-biased with language %modelling as well as to help it retain the phonetic sounds \textit{[IS SOUND a correct term?]}. 
%modelling and focus only on acoustic information in the speech.The specific details of phoneme-sequence generator is beyond the scope of this work [CITE]
% also use separate acoustic modeling independent of word sequence - not to bias wrt language modeling. 

\subsection{Phoneme sequence generator}
% We hypothesize that the phoneme sequence generated directly from ASR transcripts propagate the same errors to phoneme sequence and hence the model results may be sub-optimal. 
% Since phoneme sequence generated directly from ASR transcripts accumulates the errors in word transcripts, the resultant phoneme sequence would never to able to retain the deleted words and substitution errors in the ASR transcript. 
To address the drawback of accumulating the errors in ASR transcript by directly converting it into phoneme sequence, we generate phoneme sequence from a separate phonemic listen-attend-spell (LAS) model explained in \cite{DBLP:conf/interspeech/IriePKBRN19}. The phonemic LAS model is a sequence-to-sequence model with phoneme as its output unit. The primary objective to use an independent acoustically trained phoneme-generator is to make it non-biased with language modelling and focus only on acoustic information in the speech. We aim to retain the complementary information in the word and phonemes sequences especially for noisy speech. We use LibriSpeech data to train the phoneme model achieving the phoneme error rate (PER) of 10.2 on test-other data while 3.6 on test-clean data. 
Further specific details of phoneme LAS model is beyond the scope of this work.

\subsection{PhonemeBERT as encoder in low-resource setup}
In many practical setups, we do not always have an advantage of phoneme sequence available to us. The only resort in this case is to fine tune a vanilla word-only model on the downstream task. We demonstrate that PhonemeBERT can be used as a pre-trained encoder which when fine-tuned on the downstream task with word-only input leads to a better result than a vanilla setup (discussed in Section \ref{results_sec}). Thus, a jointly pre-trained PhonemeBERT learns phonetic-aware representations which outperform word-only representations.
% also use separate acoustic modeling independent of word sequence - not to bias wrt language modeling. 
% This step is different from the work done in \cite{DBLP:conf/sigir/FangFLR20}. Fang et al. generate phoneme sequence from the transcribed output as a post-processing step. The approach has a drawback that the errors present in the ASR transcript are propagated to the phoneme sequence. The resulted phoneme sequence would never to able to retain the deleted words and substitution errors in the ASR transcript. 
%Hence, it will not correspond to the original phoneme sequence present in the speech signal. 
% * Amazon Polly

% * ASR and Phoneme Generation

% \section{Dataset}
% \label{dataset}

% We argue that training a generic PhonemeBERT model is key to use the model for downstream tasks. 

% Amazon -> Post-processing (phonemes coming out of ASR is 'impure') - can only handle substitution errors
% Separate acoustic modeling independent of word sequence - not to bias wrt language modeling
% Pure phoneme sequences
% Discussion is beyond the scope of this paper
% Standard encoder-decoder - Recent phoneme recognition paper
% PER? - Not always less than WER

% Robust - noise
% WER - Percentage
% Beyond some level of noise,

\section{Experiments and Results}

\subsection{Dataset}
For experimental evaluation of the approach, we use three benchmark datasets as described below (Table \ref{dataset_down}). We use the exact same setup to generate the noisy version of the dataset and the corresponding ASR transcript and phoneme sequences.
\begin{itemize}
    \item SST-5: This is Stanford Sentiment Treebank dataset \cite{DBLP:conf/emnlp/SocherPWCMNP13} containing the labels in five point scale from very negative to very positive sentiment.
    \item TREC: This is a question classification dataset \cite{DBLP:conf/coling/LiR02} with coarse (6 classes) and fine-grained (50 classes) label set.
    \item ATIS: This is an  audio recordings of people making flight reservations \cite{DBLP:conf/slt/TurHH10} with intent recognition as one of the downstream tasks. We only use datapoints with single label in our experiment.
\end{itemize}

\begin{table}[t]
\footnotesize
\centering
\caption{Statistics of downstream task dataset}
\renewcommand{\arraystretch}{1.15}
\begin{tabular}{llllll}
Dataset & \#Class & \begin{tabular}[c]{@{}l@{}}Avg. \\ Length\end{tabular} & Train & Test & WER  \\ \hline
SST-5   & 5       & 17.38                                                   & 8534  & 2210 & 31.86 \\
TREC    & 6, 50   &  8.89                                                  & 5452  & 500  & 32.93 \\
ATIS    & 22       &  11.14                                                  & 4978  & 893  & 29.11
\end{tabular}
\label{dataset_down}
\end{table}

\begin{table*}[ht]
\small
\caption{Comparison of the proposed model, \textit{PhonemeBERT}, with baselines and ablation setups.}
%An ablation study without joint-modelling (A3) and using only one mode of feature for fine-tuning on the downstream task.}
\renewcommand{\arraystretch}{1.2}
\begin{tabular}{cllllll}
\multicolumn{1}{c|}{\multirow{2}{*}{\textbf{Baseline}}} & \multicolumn{1}{c}{\multirow{2}{*}{\textbf{\begin{tabular}[c]{@{}c@{}}Pre-Training \\ (Feature, Joint Loss-T/F)\end{tabular}}}} & \multicolumn{1}{c}{\multirow{2}{*}{\textbf{\begin{tabular}[c]{@{}c@{}}Task Fine- \\ Tuning Feature\end{tabular}}}} & \multicolumn{4}{c}{\textbf{Results (Accuracy/Macro-F1)}}                                                                                            \\
\multicolumn{1}{c|}{}                                   & \multicolumn{1}{c}{}                                                & \multicolumn{1}{c}{}                                                                                                         & \multicolumn{1}{c}{\textbf{SST-5}} & \multicolumn{1}{c}{\textbf{TREC-6}} & \multicolumn{1}{c}{\textbf{TREC-50}} & \multicolumn{1}{c}{\textbf{ATIS}} \\ \hline
\multicolumn{1}{c|}{Oracle}                                 & No (-, $F$)                                                            & $w_{clean}$                                                                                                                      & 56.60 / 54.03                        & 87.20 / 97.64                         & 91.80 / 86.35                          & 99.37 / 94.78                       \\ \hline
\multicolumn{1}{c|}{B1}                                 & No (-, $F$)                                                            & $w_{asr}$                                                                                                                        & 43.12 / 42.02                        & 82.60 / 80.82                         & 76.20 / 58.70                          & 94.87 / 81.67                       \\
\multicolumn{1}{c|}{B2}                                 & Yes ($w_{asr}$, $F$)                                                         & $w_{asr}$                                                                                                                        & 44.52 / 40.84                        & 85.40 / 83.53                         & 77.80 / 59.64                          & 95.05 / 81.51                       \\
\multicolumn{1}{c|}{B3}                                  & Yes ($w_{asr} + p_{asr}$, $T$)                                                    & $w_{asr} + p_{asr}$                                                                                                               & 46.24 / 38.97                        & 86.20 / 84.39                         & 80.80 / \textbf{67.45}                          & 96.75 / 80.36  \\
\multicolumn{1}{c|}{PhonemeBERT}                                  & Yes ($w_{asr} + p_{sp}$, $T$)                                                    & $w_{asr} + p_{sp}$                                                                                                               & \textbf{46.74 / 43.95}                        & \textbf{87.00 / 85.34}                         & \textbf{81.40} / 65.16                          & \textbf{97.25 / 84.15}                       \\ \hline
\multicolumn{7}{c}{\textbf{Ablation Study}}                   \\ \hline
\multicolumn{1}{c|}{A1}                                 & Yes ($w_{asr} + p_{sp}$, $T$)                                                 & $w_{asr}$                                                                                                                      & 46.70 / 41.54                        & 85.80 / 83.85                         & 79.60 / 62.77                          & 95.62 / 79.80                       \\
\multicolumn{1}{c|}{A2}                                 & Yes ($w_{asr} + p_{sp}$, $T$)                                                 & $p_{sp}$                                                                                                                        & 38.77 / 34.17                        & 83.40 / 83.94                         & 75.20 / 62.03                          & 96.25 / 79.98                       \\
\multicolumn{1}{c|}{A3}                                 & Yes ($w_{asr} + p_{sp}$, $F$)                                                & $w_{asr} + p_{sp}$                                                                                                                 & 46.10 / 43.16                        & 84.60 / 83.47                         & 79.60 / 62.05                          & 96.00 / 79.81 \\
% \multicolumn{1}{c|}{PhonemeBERT}                                  & Yes (joint: w-asr + p-asr)                                                    & w-asr + p-asr                                                   & 46.74 / \textbf{43.95}                        & \textbf{87.00 / 85.34}                         & \textbf{81.40 / 65.16}                          & \textbf{97.25 / 84.15}                       \\ \hline
\end{tabular}
\label{results}
\end{table*}

\begin{table}[]
\footnotesize
\centering
\renewcommand{\arraystretch}{1.2}
\caption{Accuracy comparison of the baseline (\textit{B1, B2}) and proposed model (\textit{P: PhonemeBERT}) at different WER range.}
\begin{tabular}{l|lll|lll}
\multicolumn{1}{c|}{\multirow{2}{*}{WER}} & \multicolumn{3}{c|}{TREC-50}                                              & \multicolumn{3}{c}{SST-5}                                              \\ \cline{2-7} 
\multicolumn{1}{c|}{}                     & \multicolumn{1}{c}{B1} & \multicolumn{1}{c}{B2} & \multicolumn{1}{c|}{P} & \multicolumn{1}{c}{B1} & \multicolumn{1}{c}{B2} & \multicolumn{1}{c}{P} \\ \hline
10-20                                     & 80.00                  & \textbf{81.43}                  & 80.00                  & 46.94                  & \textbf{51.47}                  & 50.94                 \\
20-30                                     & 81.82                  & 82.82                  & \textbf{84.85}                  & 43.62                  & 43.62                  & \textbf{46.29}                 \\
30+                                       & 69.53                  & 72.53                  & \textbf{77.52}                  & 41.04                  & 40.57                  & \textbf{43.50}                 \\ \hline
Overall                                   & 76.20                  & 77.80                  & \textbf{81.40}                  & 43.12                  & 44.52                  & \textbf{46.74}                
\end{tabular}
\label{wer}
\end{table}

\subsection{Implementation Details: Pre-training and Fine-tuning}
The PhonemeBERT model is jointly trained on ASR transcript augmented with phoneme sequence in batch size of 192 on $4 \times V100$ GPU for 50 epochs.
Once we have pre-trained PhonemeBERT, we fine-tune it for the downstream tasks with task specific cross-entropy loss.  The downstream models are fine-trained on the PhonemeBERT model for 20 epochs. The checkpoint with the best validation score is used to report the test results. For cases where a pre-defined validation set is not provided, we use 20\% of the training set as the validation set.

\subsection{Baselines and Ablations}
We define various baseline systems and ablations to compare the performance against PhonemeBERT which is jointly trained on ASR transcript (\textit{$w_{asr}$}) and corresponding generated phoneme sequence (\textit{$p_{sp}$}):
% The baseline is represented in form of $F_{mode}^{pre\mathit{-}training}$, where $F$ represents the feature used ($W$ for word sequence and $Ph$ for phoneme sequence), $mode$ represents the type of input used at the inference ($clean$ refers to original text corpus, $asr$ refers to input coming from the noisy corpus) and $pre\mathit{-}training$ is an indicator to define if pre-training is done on the generated noisy corpus ($True$).

\begin{itemize}
    % \item $W_{clean}^{pt\mathit{-}False}$
    \item \textbf{Oracle}: This baseline sets the upper limit of the results obtained by fine-tuning RoBERTa by training and testing on the original ($w_{clean}$) corpus for downstream tasks.
    % \item $W_{asr}^{pt\mathit{-}False}$: 
    % \item $M_{w_{asr}+p_{asr}}^{w_{asr}+p_{asr}}$
    
    \item \textbf{B1}: In this baseline, we fine-tune RoBERTa base model directly on the downstream tasks with task specific ASR transcripts. 
    %This method specifies an immediate fine-tuning strategy for the downstream task with a noisy data ($w_{asr}$).
    %[Fairly strong baseline as C: w-clean achieves close to SOTA]
    % \item $W_{asr}^{pt\mathit{-}True}$: 
    
    \item \textbf{B2}: We further pre-train the RoBERTa base model on our ASR transcripts from the pre-training corpus with word-only MLM task \cite{DBLP:journals/corr/abs-1907-11692}. We then fine-tune the model for each noisy-downstream task using word-only input.
    
    \item \textbf{B3}: In this setup, we pre-train a joint LM on word and phoneme sequence along with fine tuning it on downstream tasks with both information fed to the model. The phoneme sequence (\textit{p$_{asr}$}) in this case is generated directly from the text using Phonemizer\footnote{\url{https://github.com/bootphon/phonemizer}}  tool.
    
    \item \textbf{A1}: This ablation model refers to using PhonemeBERT as word-only encoder in the low-resource setup where only ASR transcript is available for downstream tasks.
    
    \item \textbf{A2}: This setup is similar to A1 except that A2 uses only phoneme sequence for the downstream tasks.
    
    \item \textbf{A3}: In this setup, we train a PhonemeBERT-like model without a joint loss function i.e, with separate word-only and phoneme-only loss.
    
    % \item $W_{asr}^{pt} + Ph_{asr}^{pt} $
\end{itemize}

% The proposed model is denoted as $J_{asr}^{p\mathit{-}True}$, where we jointly model both ASR transcript and phoneme sequence.

\subsection{Results and Analysis}
\label{results_sec}
The comparative results of the proposed approach against various baselines are presented in Table \ref{results}. \textit{PhonemeBERT} trained with joint modelling framework of ASR transcript and phoneme sequences outperforms other models across all datasets. The proposed model comprehensively beats the model pre-trained on only ASR transcripts (\textit{B2}) by upto 3.6\% in TREC-50 on accuracy. This shows that a language model trained on noisy ASR transcript alone is not robust to ASR errors as compared to the one trained with both ASR transcripts and phoneme sequences. Additionally, PhonemeBERT also outperforms \textit{B3} which justifies our hypothesis that a model trained on phoneme sequence generated directly over ASR transcripts yields sub-optimal results.

%The proposed model achieves better accuracies by up to 3.4\% compared to the model pre-trained with only ASR transcript (\textit{B3}). 
Furthermore, in a low-resource setup (\textit{A1}) where phoneme is not available for the downstream task, the model performs better than a word-only model (\textit{B2}) fine-tuned on the downstream task.
% the baseline \textit{B2} performs %worse than a model pre-trained with joint modelling setup but fine-tuned on downstream tasks using only ASR transcripts as input (A1). 
This result justifies our hypothesis that joint modelling helps with learning a phonetic aware language model that is more powerful in noisy conditions.
%\textit{[HAVEN'T USED PHONETIC-AWARE REPRESENTATIONS TERMINOLOGY ANYWHERE EARLIER, either rephrase it here or mention it in introduction]}. 
% In the absence of phoneme sequence during inference, the jointly pre-trained model can be used for downstream task fine-tuning with only ASR transcripts as input feature.
We also evaluate the setup using only phoneme sequence for downstream fine-tuning (\textit{A2}). The results show a drop in evaluation scores for all but one dataset. We note a sharp decline in results for SST-5 and TREC-50 dataset compared to other results. This can be explained by the fact that both of these two tasks need understanding of semantics in the text than syntactic information. Phoneme sequences do not capture the semantic information, rather they are syntactic representation of the utterances and hence there is a loss of performance phoneme-only downstream setup.
Finally, to study the impact of joint-modelling framework, we pre-train a model with independent word-MLM and phoneme-MLM tasks \textit{A3}, where each task sees the context from respective sequence only. PhonemeBERT performs better than the model trained on independent MLM tasks with big gains in F1-score on TREC-6, TREC-50 and ATIS datasets.

% We also present a qualitative analysis demonstrating the scenarios effectively captured by the proposed model against the word-only baseline 
% In Table \ref{qualitative}, we present an example where the ASR transcription is completely erroneous but the presence of \textit{why} in the phoneme sequence helps in the correct classification. 
%Similarly, in the second example, the presence of phoneme sequence helps to correct some of the ASR errors to arrive at the correct output. 
\begin{table}[]
\footnotesize
\centering
\renewcommand{\arraystretch}{1.2}
\caption{Results on Observe.AI's sentiment classification task}
\label{oai_sentiment}
\begin{tabular}{c|ccc}
\textbf{Model}       & \textbf{Precison} & \textbf{Recall} & \textbf{F1}    \\ \hline
B1                   & 60.32             & 79.29           & 68.51          \\
B2                   & 65.06             & 76.54           & 70.33          \\
A1                   & 67.28             & 77.18           & 71.79          \\
\textbf{PhonemeBERT} & \textbf{69.16}    & \textbf{76.16}  & \textbf{72.49}
\end{tabular}
\end{table}

In a separate analysis, we categorize the test set of \textit{TREC-50} and \textit{SST-5} into buckets based on the WER score of each instance and compare the performance of the trained models (Table \ref{wer}) on each bucket. We note that for a lower WER ($10\mathit{-}20$), the baseline pre-trained on only ASR transcripts performs best. However, at higher WER ($>=20$), the performance of the two models (\textit{B1, B2}) decline steeply, while PhonemeBERT outperforms the word-only models and is less susceptible to the noise. We speculate that since pre-training data has an average WER greater than 20, PhonemeBERT learns to better model the data belonging to higher WERs while degrading a bit on lower WER ranges. Nevertheless, this reinforces the fact that we need a better modelling ability to address the robustness to ASR errors. For WER $>= 30$, PhonemeBERT outperforms the next-best model (\textit{B2}) by 4.99\% and 2.93\% in accuracy for TREC-50 and SST-5 respectively.
We additionally compare the proposed model on Observe.AI's sentiment classification dataset (Table \ref{oai_sentiment}). This dataset comprise of call center voice conversations with labels as negative or positive sentiment displayed by the customer on a call. The overall training and test size of the sample is 10492 and 3198 respectively with an average WER of 24.76. The results show that the joint LM setup outperforms word-only baseline \textit{B2} by 2.16 points. PhonemeBERT when used in the low-resource setup (\textit{A1}) outperforms word-only setup (\textit{B2}) by 1.46 F1 score. This demonstrates the applicability of the proposed method in the practical setups too.

% Please add the following required packages to your document preamble:
% \usepackage{multirow}
% Please add the following required packages to your document preamble:
% \usepackage{multirow}

\section{Conclusions and Future Work}
In this work, we propose a joint modelling of phoneme sequence and ASR transcript to build a language model robust to ASR errors. We demonstrate that the proposed joint learning setup performs better than any strong baselines in downstream tasks. Through ablation study, we show that the proposed setup can also work in a low resource setup, still producing better results than model pre-trained with only ASR transcript. Our analysis suggests that the method is significantly better at higher WER ranges. In future, we would like to extend the work to tasks such as entity recognition and question answering for the noisy data. Additionally, we would also like to explore the possibility of using a larger corpus for the pre-training step.

\bibliographystyle{IEEEtran}

\bibliography{mybib}

\end{document}